\documentclass[aps,pre,superscriptaddress,groupedaddress]{revtex4}  
\usepackage{graphicx}   
\usepackage{dcolumn}    
\usepackage{bm}        
\usepackage{amssymb}   
\usepackage{amsmath}
\usepackage{xcolor} 
\begin{document}
\title{Short-Range Migration Can Alter Evolutionary Dynamics in Solid Tumors}
\author{Youness Azimzade}
\email[Email:~]{y_azimzade@ut.ac.ir}
\affiliation{Department of Physics, University of Tehran, Tehran 14395-547, Iran}
\author{Abbas Ali Saberi  } 
\email[Email:~]{ab.saberi@ut.ac.ir}  
\affiliation{Department of Physics, University of Tehran, Tehran 14395-547, Iran}
\affiliation{Institut f\"ur Theoretische Physik, Universitat zu K\"oln, Z\"ulpicher Strasse 77, 50937 K\"oln, Germany} 
\date{\today}
\begin{abstract}
 Here, we investigate how competition in the Eden model is affected by short range dispersal and the requirement that site updates occur only after several updates of the same site have been attempted previously. The latter models the effect of tissue or media resistance to invasion.  We found that the existence of tissue intensifies \textit{Natural Selection} and de-accelerating \textit{Genetic Drift}, both to a limited extent.  
More interestingly, our results show that short-range migration can eliminate  Genetic   demixing  and conceal Natural Selection.   
\end{abstract} 
\maketitle
 
\section{Introduction} 
 Tumors are known to be composed of different sub-clones  \cite{fidler1978tumor} which may carry different properties and interact with each others in various ways \cite{ yachida2010distant, cleary2014tumour}. Study of interaction between these sub-clones, also known as \textit{Evolutionary Dynamics},  has gained increasing attention in the last decade, especially due to possible treatment applications \cite{cunningham2011evolutionary, gatenby2019eradicating}. Tumors can be considered  as an ecosystem of interacting sub-clones \cite{mcgranahan2017clonal,   maley2017classifying}, but as spatially structured populations \cite{de2014spatial, gonzalez2002metapopulation, campbell2010patterns, gerlinger2014genomic, yachida2010distant, navin2011tumor}, they may not follow theories for well-mixed populations \cite{korolev2010genetic}. Unfortunately, analytical understanding of structured populations is limited to specific structures \cite{frean2013effect} which necessitates development of different models to understand tumor evolution \cite{komarova2014spatial, waclaw2015spatial, nanda2018genotype, van2019spatiotemporal, west2019tissue, ardavseva2019evolutionary}. 
 
 Between tumor cells, some have higher fitness due to driver mutations \cite{martincorena2017universal}.  Natural Selection implies that between coexisting populations, including tumors \cite{michor2004dynamics, anderson2006tumor}, the one with higher fitness can take over the whole environment \cite{fisher1937wave, fisher1999genetical}. 
 Tumor cells face the crowded microenvironment as a physical barrier \cite{wolf2013physical}  that can be displaced and deformed \cite{pirentis2015remodeling}.
 Such a barrier plays the role of environmental stress in bacterial populations \cite{ben1994generic}. Environmental stress affects evolutionary dynamics in tumors \cite{anderson2006tumor, azimzade2018role}. Likewise, limits on available space or the geometry of host tissue is believed to affect tumor evolution \cite{west2019tissue, pavlogiannis2018construction}. However, the effect of this "removable spatial barrier" on the evolution of neutral populations remained to be explored.  
 On the other hand, tumors may follow \textit{Neutral} evolution \cite{graham2017measuring} in which evolution of populations is random and not based on fitness differences \cite{turajlic2019resolving}.

 Between two competitive populations with the same fitness, one of them may randomly become dominant, a process which is known as \textit{Genetic Drift} \cite{allendorf1986genetic} and has been studied using Moran model \cite{moran1958random}. While spatial Moran models have been developed for specific geometries \cite{komarova2006spatial, durrett2015spatial}, the role of other structures and space-associated processes such as migration on genetic drift remained to be studied. 
 
Migration generally is referred to as long-range dislocations which play a central role in different areas of tumor biology \cite{van2011initial,  azimzade2019regulation}. Short-range migrations, sometimes referred to as dispersal \cite{waclaw2015spatial}, also are prevalent in tumors and responsible for the infiltrative edge of the tumor where some cells are slightly detached from invasion front in tumors \cite{koelzer2014tumor} or model systems \cite{ray2018dynamics}. Migrations directly affect entities encounters \cite{azimzade2017search} and respectively should regulate population dynamics. The effect of migration on evolutionary dynamics of populations living on graphs has been studied \cite{krieger2017effects} but their possible effect on evolutionary dynamics of more realistic structures remained to be addressed.  
 
In this letter, along with a model on cooperative populations in the presence of removable spatial barrier \cite{ben1994generic}, we develop a border driven growth model to study evolutionary dynamics in  cellular aggregates.  More specifically, we study how the presence of environment as a removable spatial barrier and short-range migration affect genetic drift (natural selection) for two (non-)identical competitive populations. We then interpret our findings in the context of tumor evolutionary dynamics. 
 
\section{Model} 
An interesting approach to study population dynamics is the stepping stone model in which populations are defined on a spatial lattice with each unit containing well-mixed population.  In this approach, species undertake different processes such as mutation and migration \cite{korolev2010genetic}. If we consider only one species for each unit, the stepping stone model becomes Eden model \cite{eden1961two}. Interestingly, the Eden model happened to perfectly predict \cite{saito1995critical} the dynamics of competitive bacterial populations \cite{hallatschek2007genetic}. Since then, Eden model has been used widely for further research in population dynamics studies \cite{dai2014spatial, reiter2014range, lavrentovich2013radial, greulich2012interplay, lavrentovich2014asymmetric, chu2019evolution}.
 
Similar to two species Eden model \cite{saito1995critical}, we consider two populations, (A) and (B), which are living on a 2D or 3D lattice and each unit can be empty or occupied by only one species. Then we study these populations in two different scenarios, identical case and non-identical case. The identical case represents two similar populations with the same fitness where their evolution is based on neutral evolution. The non-identical case represents two populations with different fitness. 
 
To introduce temporal evolution, we randomly choose a unit. If the unit is occupied, in identical scenario both populations will be updated according to the same rules as following:  $i$) cell decide to duplicate into an empty neighbor unit with the probability of $R$ and would do so if that unit is occupiable. We fixed $R=0.1$ in the whole paper.  $ii$) If the cell does not duplicate, it decides to migrate to an empty neighbor unit with a probability of $d$ and does so if the selected unit is occupiable. $iii$) Any unit would be occupiable after $N$ times being selected for occupation through migration or duplication.  $N$ represents the crowdedness of the environment and its effect is similar to the noise reduction process in previous studies \cite{wolf1987noise, kertesz1988noise}. Hitting represents cells effort to deform and displace host tissues components and pave their way for migration/duplication. If we set $d=N=0$, our model would be the Eden model with two species \cite{saito1995critical}.
 
Non-identical scenario considers the fitness difference between cells.  Fitness difference is not significant and typically can be considered about 10 percent \cite{heide2018reply}. To incorporate a typical fitness difference, we consider (B) to have a slight advantage in reproduction rate as $R_B=1.1 R_A=0.11$.
 
 \begin{figure} [!h]
 	\centering
 	\includegraphics[width=0.5\linewidth]{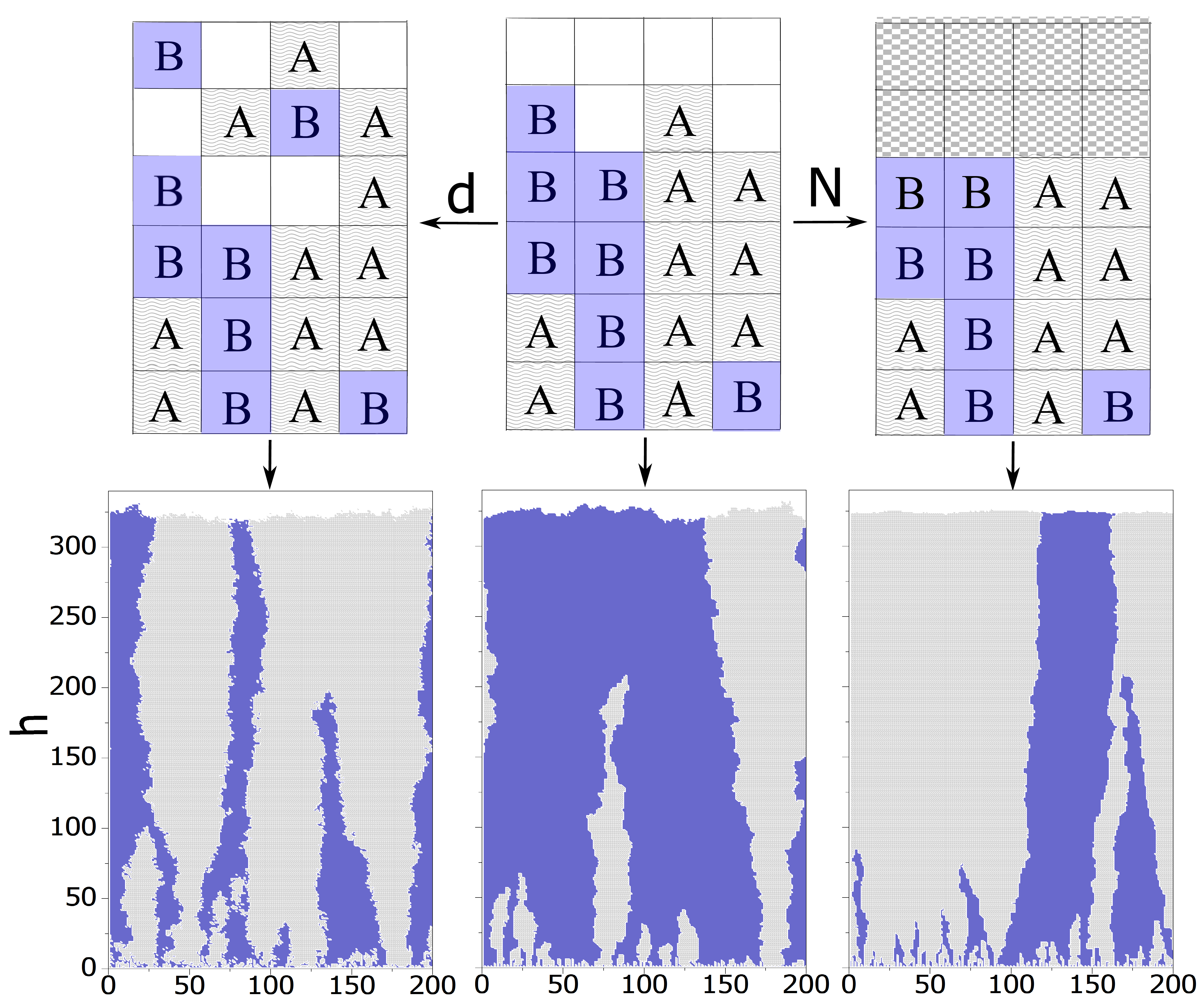}      
 	\caption{ Structure of the model where we took the Eden model and added two modifications, short-range migration (depicted by the probability of $d$) and removable spatial barrier (depicted by a density of $N$). When we add short-range migration, the infiltrative edge appears at the invasion front. Removable spatial barrier smooths invasion front. }
 	\label{FIG1}
 \end{figure} 
 
 \section{2D Results}
 We define a $600\times800$ $2D$ lattice on which cells can proliferate and migrate. We also consider periodic boundary conditions in $x$ direction. As the initial condition, in order to represent a well-mixed population, we put (A) and (B) in neighbor units in the first five rows. We start with the simplest case, $d=N=0$. Similar to experimental results for bacterial colonies \cite{hallatschek2007genetic} and previous simulations \cite{ali2013scale, kuhr2011range}, after a few generations, segregation happens (FIG \ref{FIG1}). One relevant parameter to quantify this behavior is the number of units in the interface of two populations, $n_{AB}$ \cite{saito1995critical, korolev2010genetic}. To obtain this quantity, we let the two populations grow and occupy 90 percent of the environment. Then, at each row, we count the number of (A) cells that are in contact with one or more (B) cells in their four nearest neighbors. This parameter gives us a measure for diversity. For first few rows where (A) and (B) pairs are juxtaposed, each (A) is contact with at least one (B) and the $n_{AB}$ is equal to the number of (A) and $L/2$, respectively.  Since  $n_{AB}$ exhibits fluctuations, to have a better measure, we will obtain its expectation value,  $\langle n_{AB} \rangle $, by averaging over 100 realizations. 
 
 As we move towards higher rows, more and more (A) cells lose their contact with (B)  cells and  $\langle n_{AB} \rangle $  falls down.  From Stepping stone model we anticipate having $ \langle n_{AB} \rangle \propto h^{-\beta}$ with $\beta=1/2$ where $h$ stands for height in $y$ direction. However, because of surface irregularities, we are observing $\beta=0.67\pm 0.02$.  This result is in perfect agreement with experimental results for bacterial colonies \cite{hallatschek2007genetic} and numerous on/off-lattice models \cite{saito1995critical, ali2013scale, kuhr2011range}. 
 
\begin{figure} [!h]
 	\centering
 	\includegraphics[width=0.5\linewidth]{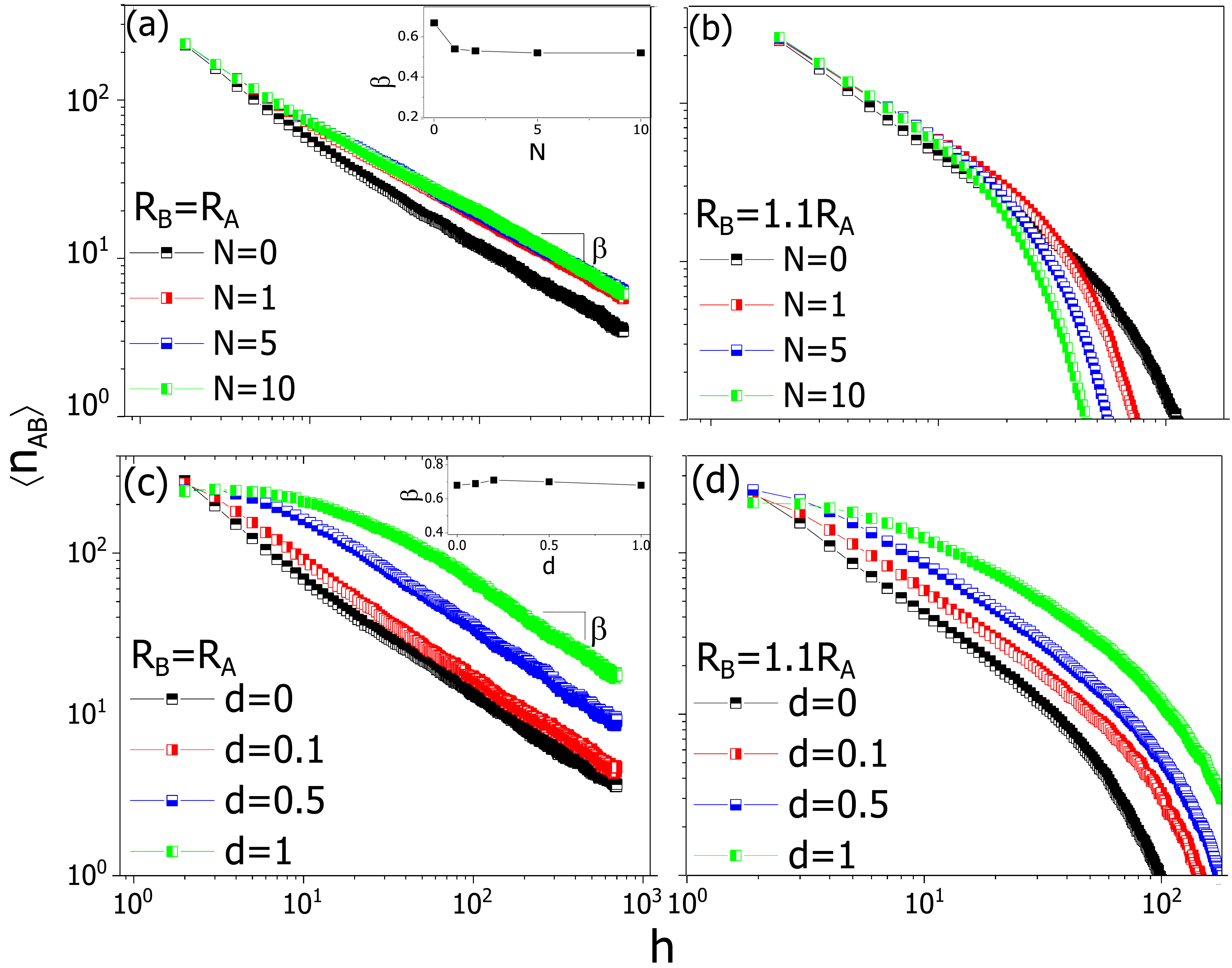}     
 	\caption{ (a) $\langle n_{AB} \rangle$ versus $h$ for identical populations and different values of $N$ where it changes the dynamics of $\langle n_{AB} \rangle$. Inset: $\beta $  versus $N$. As such, existence of host tissue as a removable spatial barrier slows down neutral evolution by de-accelerating genetic drift. (b) $\langle n_{AB} \rangle$ versus $h$ for non-identical populations where $N$ decrease $\langle n_{AB}\rangle$. As such, environment amplifies  natural selection. (c) Effect of short-range migration on $\langle n_{AB} \rangle$. Migration increase $\langle n_{AB} \rangle$ but the dynamics of $\langle n_{AB} \rangle$ remains the same with  $\beta\sim0.67$. This shows that in 2D, short-range migration favors coexistence of Neutral populations but it does not affect the dynamics of genetic drift. (d) For non-identical populations, short-range migration increases $\langle n_{AB} \rangle$ and favors co-existence of populations.}
 	\label{FIG2}
 \end{figure}

 We then consider the removable spatial barrier by setting $N\ne0$. When cells are facing a barrier to overcome, the growth becomes slower. More interestingly, the front becomes more smooth and for $N=10$, we have $\beta=0.52 \pm 0.02$ (FIG \ref{FIG2} (a)). Smooth front means that we are close to Row-by-Row growth which follows the stepping stone model with $\beta=1/2$ \cite{korolev2010genetic}. For non-identical populations, when the front goes smoother, $\langle n_{AB} \rangle$ falls faster versus $h$ (FIG \ref{FIG2} (b)).  As such, environment cancels out invasion front fluctuations and makes the beneath population to follow the stepping stone model (standard Moran model), where we have $\beta= 1/2$ for identical population and Darwinian evolution is more effective between populations with different fitness.  
 
 When we allow cells to migrate by setting non-zero $d$, the infiltrative edge emerges. It should be noted that migration effectively happens only at invasion front and  cells in the bulk can not migrate because there is no free neighbor unit there. Accordingly, in addition to few properly segregated areas, smaller islands appear (FIG \ref{FIG1}). This behavior leads to an increase in $\langle n_{AB} \rangle$. In long time, however, dynamics versus $h$ remains unchanged with the same $\beta$. As such, genetic drift in the presence of short-range migration which is associated with an infiltrative edge still follows the standard Eden model with $\beta\sim 0.67\pm0.03$ (FIG \ref{FIG2} (c)). For non-identical populations, short-range migration appears to prolong coexistence of two populations (FIG \ref{FIG2} (d)). Thus, short-range migration weakens natural selection.  

 The rational behind this behavior  is that when $d=0$, the velocity of the expansion is
 proportional to $R$. When $d \ne 0$, both $d$ and $R$ contribute in invasion. In the continuum
 limit, the velocity is $2 \sqrt{Rd} $. This change in the dependence on $R$, reduces the difference in the velocities of the species. Therefore, the effective fitness differences change with $d$. 
 
\begin{figure} [!h]
 	\centering   
 	\includegraphics[width=0.5\linewidth]{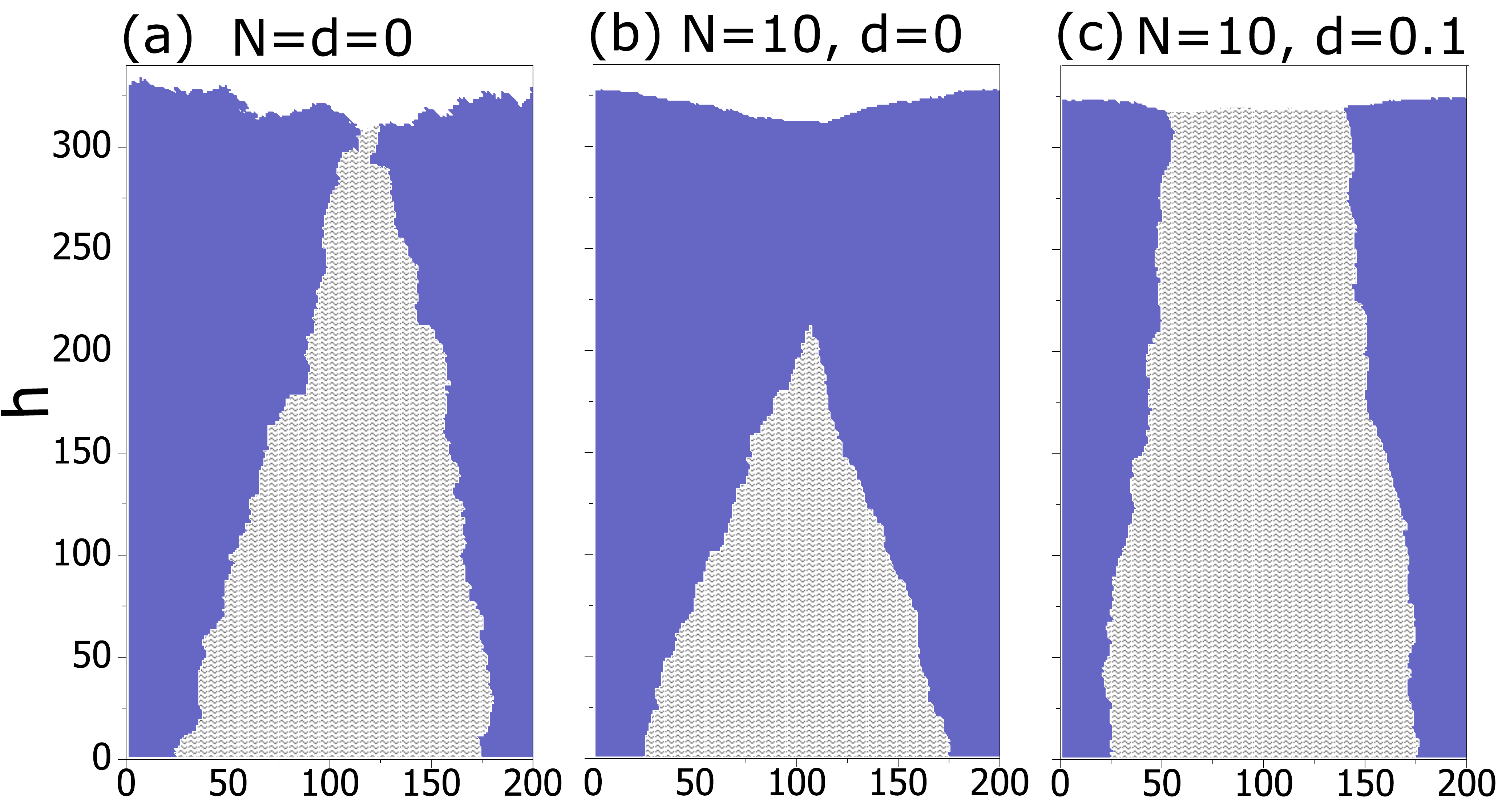}   
 	\caption{ (a) Eden model with $R_B=1.1R_A$ where (B) finally becomes dominant. (b) Adding physical barrier decreases the survival chance for (A) and (B) becomes dominant earlier. (c) Adding  short-range migration ($d=0.1$) increases survival chance for (A) and postpones fixation of (B).}
 	\label{FIG3}
 \end{figure}

 Our model prediction about the role of short-range migration is in line with recent findings on the effect of physical parameters on natural selection in bacterial colonies \cite{giometto2018physical, kayser2019collective}. Based on one of these studies \cite{ kayser2019collective}, the collective migration of cells conceals fitness differences. However, our result predicts that even non-collective migration can increase the competition chance for the population with smaller fitness.  FIG \ref{FIG3} depicts how such migration changes the efficiency of natural selection. In standard Eden model with $  R_B=1.1 R_A  $, (B) finally takes over the entire invasion front (FIG \ref{FIG3} (a)). Adding removable physical barrier accelerates the fixation of (B) (FIG \ref{FIG3} (b)) and adding short-range migration weakens natural selection (FIG\ref{FIG3}(c)).   
 
Our results for the 2D environment shows that population dynamics is not only a function of fitness differences. Instead, properties such as short-range migration, which may seem irrelevant, interfere with natural selection and increase the survival chance for populations with smaller fitness, conceptually in the line with previous results \cite{tilman1994competition}.
Additionally, the presence of host tissue as "removable spatial barrier" accelerates natural selection, similar to other environmental stresses. These findings can be experimentally analyzed using cancer cells or bacterial colonies on 2D substrates \cite{hallatschek2007genetic, huergo2014dynamic}. However, to understand the process in more realistic settings, we develop 3D models.   
 
\section{3D Results} 
We define a $50\times50\times500$ 3D lattice on which cells can proliferate and migrate. As the initial condition, similar to 2D, we put (A) and (B) in neighbor units to imitate   a well-mixed population for first five  $xy$ plains and then let them evolve according to the aforementioned rules. In both $x$ and $y$ directions, we consider periodic boundary conditions.
We let the system to evolve until 90 percent of the environment get occupied. For standard Eden model with identical populations, segregation happens and two populations occupy different areas only after a few generations.
Here we count $n_{AB}$ as the number of (A) cells which have a nearest neighbor unit occupied by a (B) cell. Expectation value for number of border units falls as  $\langle n_{AB} \rangle \propto h^{-\beta}$ with $\beta=0.33\pm0.02$. 
 For non-identical populations, we anticipate (B) to be fixed and $\langle n_{AB} \rangle$ to reach to zero in long time.  
 
\begin{figure} [!h] 
\includegraphics[width=0.5\linewidth]{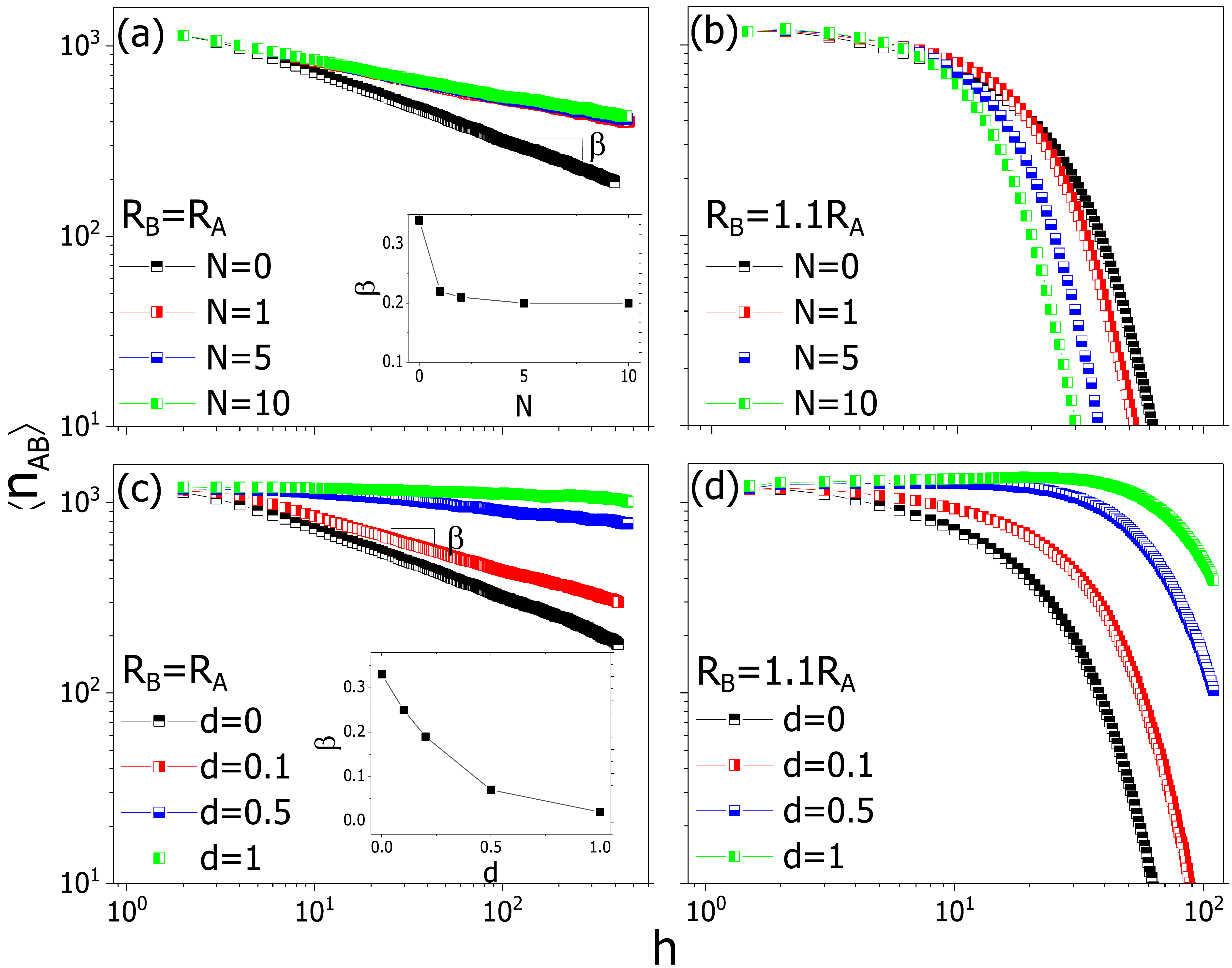}  
\caption{(a) $\langle n_{AB} \rangle$ versus $h$ in presence of removable spatial barriers for identical populations. Inset: Stress hinders genetic  demixing  by decreasing $\beta $ from $0.33\pm 0.02$ to $0.20\pm 0.01$.  (b)  $\langle n_{AB} \rangle$ versus $h$ for non-identical populations where  $N$  accelerates Darwinian evolution by amplifying natural selection. 
(c) $\langle n_{AB} \rangle$ versus $H$ for identical  populations in presence of short-range migration. Inset: Short-range migration leads to $\beta  \sim 0$ which implies zero genetic  demixing  in 3D. (d) $\langle n_{AB} \rangle$ versus $h$ for $R_B=1.1R_A$, where short-range migration hinders natural selection and even conceals it for large $d$.}
\label{FIG4}
\end{figure}

 When we add environmental barrier for identical populations, genetic drift decreases (FIG \ref{FIG4} (a)). For larger numbers of $N$, we have $\beta=0.20\pm0.01$. For non-identical populations, $N$ accelerates the fixation and decreases $\langle n_{AB} \rangle$ (FIG \ref{FIG4} (b)). The effect of $N$  on  $\beta$ is similar to 2D case.  
 
 Interestingly, considering $d\ne0$ leads to unexpected results in 3D.  For identical populations, short-range migration starts to suppress $\beta$ and for $d=1$ we have $\beta=0.02$ (FIG \ref{FIG4} (c)). As such, short-range migration can lead to the eradication of spatial segregation by making populations well-mixed and eliminate genetic demixing. Note that in 2D, migration increased $\langle n_{AB} \rangle$ but $\beta$ remained the same.  
 As FIG \ref{FIG4} (d) shows, for non-identical populations, adding short-range migration increases $\langle n_{AB} \rangle$ which means that in spite of being well mixed, natural selection is weaker between these populations. In fact, for $d =1$, natural selection remains quite ineffective for a considerably long time.  
 
Our model predicts that the existence of removable spatial barrier acts as environmental stress similarly both in 2D and 3D. On the other hand, the effect of short-range migration is different for 2D and 3D. In fact, one may consider its effect in 2D as a minor effect since it does not affect $\beta$, yet, in 3D it changes the dynamics profoundly, by eliminating genetic  demixing  and hindering natural selection for seemingly well-mixed populations.

\section{Re-analyzing Natural Selection} 
\begin{figure} [!h] 
\centering  
\includegraphics[width=0.5\linewidth]{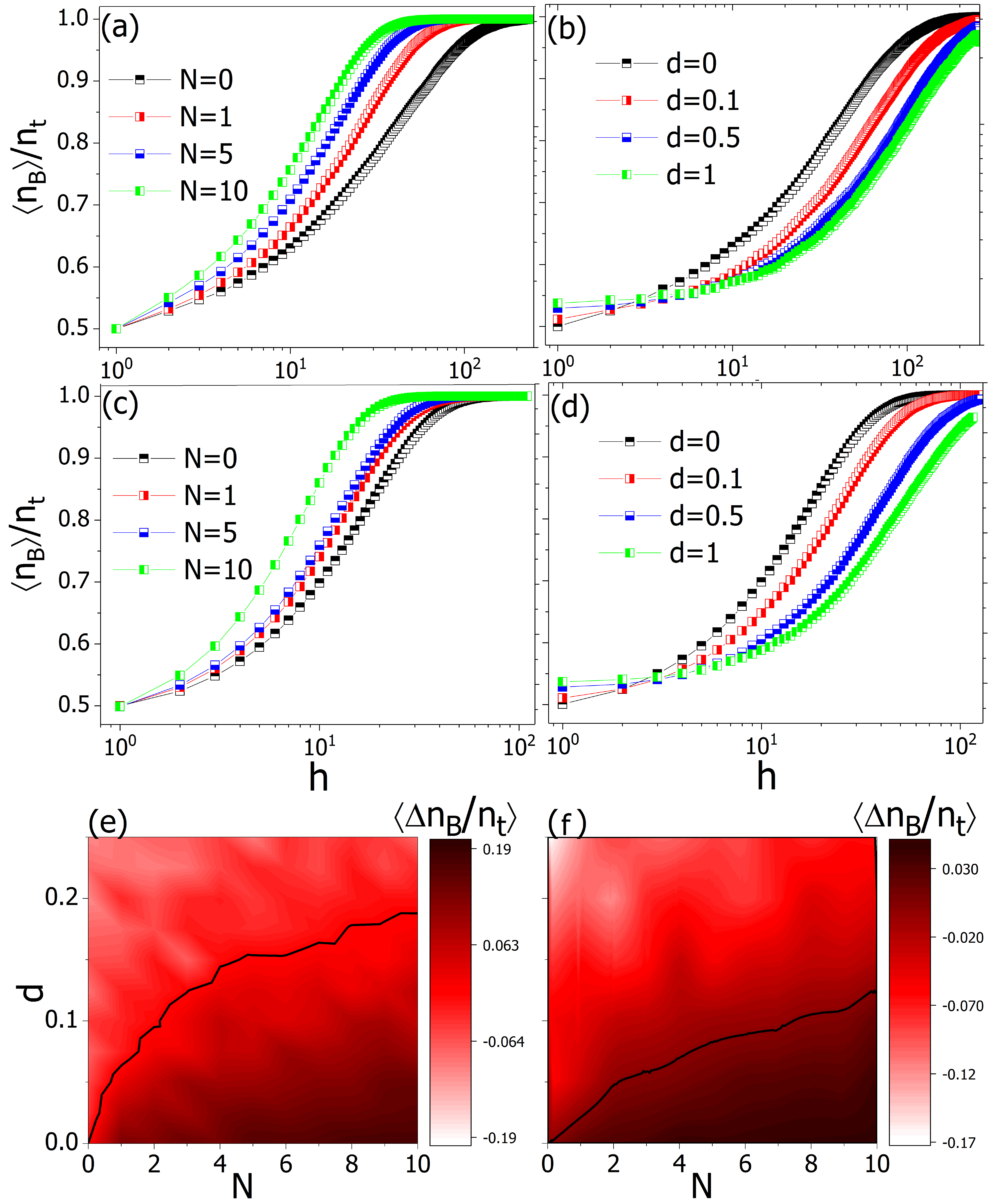}        
\caption{  (a)Behavior of $ \langle n_{ B}\rangle/n_t$ versus $h$ in 2D for different values of $N$. $N$ increases $ \langle n_{ B} \rangle /n_t$. $ \langle n_{ B} \rangle /n_t$  versus $h$ for different values of $d$. $d$ decreases $ \langle n_{ B} \rangle /n_t$  and de-accelerates natural selection. Behavior of $ \langle n_{ B} \rangle /n_t$  versus $h$ for different values of $N$ (c) and $d$ (d) at 3D. These results confirm the role of $N$ and $d$ in accelerating and weakening natural selection, respectively. Behavior of $ \langle\Delta n_{ B}(N,d)\rangle  /n_{t}=\langle n_{ B}(N,d)-n_{ B}(0,0)\rangle /n_{t}$ for different values of $N$ and $d$ in 2D (e) and 3D (f). Solid line shows $\langle\Delta n_{B}\rangle /n_{t}=0$.  These results show that interplay between $N$ and $d$  determines efficiency of natural selection.} 
\label{FIG5}
\end{figure}

 Here we try to analyze natural selection more carefully. To do so, we directly analyze the number of (B) cells which have higher fitness. Due to fitness advantage, (B) finally becomes dominant but we will study how fast the the domination happens. For both 2D and 3D configuration we study evolution of $N_{B}$. FIG \ref{FIG5} (a) shows that $N$ increases $n_B$ which confirms that environment intensifies natural selection. Short-range migration, as FIG \ref{FIG5} (b) shows, decreases $n_B$ which is in line with previous results (FIG \ref{FIG2} (d)) where we  found it to de-accelerate natural selection.  Analysis of  $n_B$ for 3D (FIGs \ref{FIG5} (c) and(d)) confirms our conclusion based on $n_{AB}$. 
 As FIGs \ref{FIG2} and \ref{FIG4} show, for non-identical scenario, effect of short-range migration and physical barrier cancel each other. To find the winner between these two competitive parameters for different values of $N$ and $d$. We set $N=d=0$ as our reference frame and study the variations in $ \langle \Delta n_{B} (N,d)   \rangle /n_{t} = \langle n_{B}(N,d)-n_{B}(0,0)\rangle / n_{t}$, where $n_{t}=n_{A}+n_{B}$, to find how interplay between parameters affects the fraction of (B) cells and natural selection. Positive (negative) values of $\langle \Delta n_{B}(N,d)\rangle  /n_{t}$ show more (less) effective natural selection.  
 FIG \ref{FIG5} confirms the role of $N$ and $d$ on  natural selection and shows that interplay between the two parameter affects the efficiency of natural selection. 
 
\section{Discussion}  
\begin{figure} [!h]
\centering   
\includegraphics[width=0.5\linewidth]{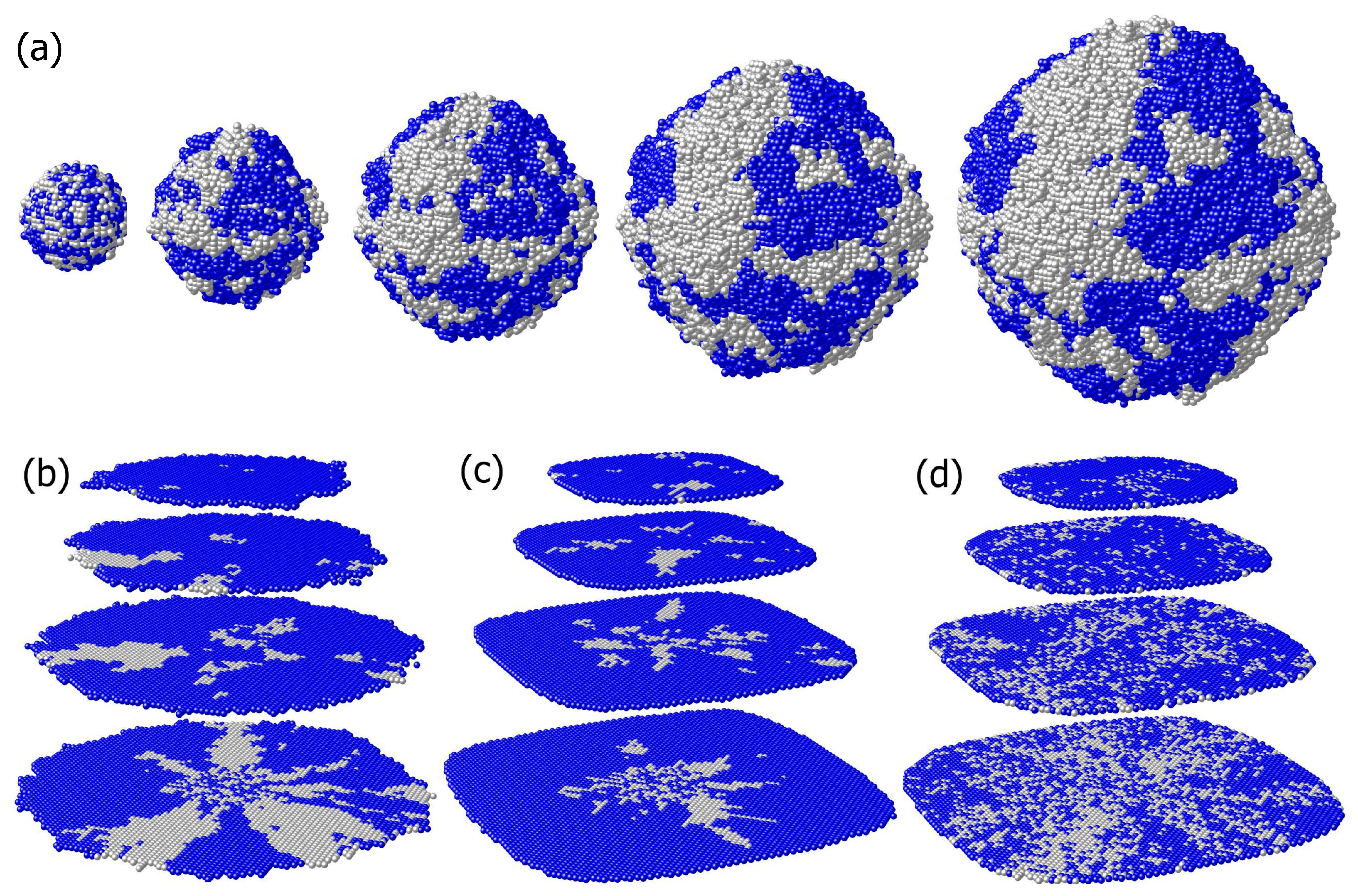}    
\caption{(a) Growth of 3000 well-mixed  (A) and (B) cells in the identical scenario in the Eden model. As expected, spatial segregation appears during growth. (b) Few slices of the grown tumor, containing two populations in which $R_B=1.1 R_A$,  for 3D Eden model. (A) has a chance to survive and appear at the surface of the tumor but in longer times the chance reaches to zero. (c)  Growth in the presence of removable barrier where fixation is accelerated and respectively the chance for (A)  to remain active at the surface decreases. (d) When we add short-range migration, survival chance for (A) increases significantly and it appears at the tumor border and actively participates in duplication and growth. The short-range migration changes the dynamics of growth and leads to deviation of natural selection and respectively Darwinian evolution.}
\label{FIG6}
\end{figure}

 As mentioned earlier, our findings can be applied to different cellular aggregates such  as bacterial colonies and tumors. However, we use this model to address the evolutionary dynamics in solid tumors. 
To show our findings in the context of tumor evolution, we use our 3D model to imitate tumor growth. We consider 3000   cells of (A) and (B) randomly positioned in the center of the 3D medium and let them grow according to the aforementioned rules. For two similar populations,  both of them will appear at the surface and probable fixation is unlikely to happen due to the geometry of the tumor ( see FIG \ref{FIG6}(a)). Then we consider the non-identical scenario with  $R_B=1.1 R_A$. Based on the neutral theory, (A) should actively participate in the growth and should be present at the border of the tumor, in spite of having lower fitness. Based on Darwinian evolution, (B) is dominant and it will take over the whole tumor and will be the only population active in the border.  
 
 For the Eden model, as FIG \ref{FIG6}(b) shows, (A) has a chance to appear at the border but  finally it will be captured inside the tumor as the tumor becomes larger. When we add environmental barrier, the situation for (A) becomes worse and it will be run out of border faster (FIG \ref{FIG6}(c)). However, when we add short-range migration by keeping $N=10$, the situation becomes completely different. First, two populations, as FIG \ref{FIG6}(d) shows, became well-mixed.  Then, in spite of being mixed, natural selection is not effective as before and (A) remains active in the border and participates in tumor evolution. Interestingly, while the dynamics of sub-clones has changed, the morphology of the tumor remains almost the same. 
 
 Recent advancements allowed to spatially track different clones in the tumor \cite{lamprecht2017multicolor, van2019spatiotemporal}, paving a way to understand tumor evolutionary dynamics. However, little is known about parameters regulating the evolution of populations,  due to lack of theoretical models for 3D structured populations.  Our model makes two main predictions that can be checked experimentally using recent system models. First, it predicts that if growth happens in the medium with a removable spatial barrier, such as a gelled medium, selection will be stronger and Darwinian evolution would be more effective. The second prediction is that if cells in system model have the ability to migrate, even for a limited length, the spatial structure of sub-clones would be entirely different, while the geometry of the whole tumor may remain almost the same. In those seemingly well-mixed populations, natural selection and genetic drift will be much weaker.

 Spatial models have a long and rich history in cancer modeling \cite{williams1972stochastic}. The effect of space and spatial heterogeneity has been considered in numerous studies \cite{anderson2006tumor, ardavseva2019evolutionary, gallaher2018spatial} but the results normally favor natural selection. More recently, we found that environmental disorder can change the composition of the invading population, possibly leading to changes in evolutionary dynamics of tumors  \cite{Azimzade2019disorder}. 
 Here, our results reveal  the importance of being discrete and spatial \cite{durrett1994importance} by demonstrating that spatial models can provide evidence for deviation from Darwinian evolution.
 
 \section{Conclusion}
 Motivated by the convenient role of clonal interactions and environmental stresses in tumors, we developed a model to study how short-range migrations and the presence of the environment as a spatial barrier together regulate   evolutionary dynamics in cellular aggregates and applied our findings to solid tumors. We found that crowdedness of environment affects tumor evolution as environmental stress. We then found that short-range migration can conceal genetic  demixing  and natural selection. 
 Our results provide evidence for the possibility of violation of natural selection in well-mixed populations.
 These findings can be verified, thank recent experimental advancements.

\end{document}